\documentclass[12pt,prc,showpacs,preprintnumbers,unsortedaddress,amsmath,amssymb,floatfix]{revtex4}

\input epsf.sty
\usepackage{graphicx}
\newcommand{\bff}[1]{{\mbox{\boldmath $#1$}}}
\usepackage{dcolumn}
\usepackage{bm}
\usepackage{amsmath}
\usepackage{longtable}

\begin{document}
\title{Effect of Resonant Continuum on Pairing Correlations in the Relativistic Approach}

\author{Li-Gang Cao}

 \affiliation{Institute of High
Energy Physics, Chinese Academy of Sciences, Beijing 100039}

\author{Zhong-Yu Ma}

\thanks{also Center
of Theoretical Nuclear Physics, National Laboratory of Heavy Ion
Accelerator of Lanzhou, Lanzhou 730000, P. R. of China and
Institute of Theoretical Physics, Beijing 100080, P.R. of China}

 \affiliation{ CCAST, Beijing 100080, P.R. of China}

\affiliation{China Institute of Atomic Energy, Beijing 102413,
P.R. of China}

\date{\today}

\begin{abstract}
A proper treatment of the resonant continuum is to take account of
not only the energy of the resonant state, but also its width. The
effect of resonant states on pairing correlations is presented in
the framework of the relativistic mean field theory plus
Bardeen-Cooper-Schrieffer (BCS) approximation with a constant
pairing strength. The study is performed in an effective
Lagrangian with the parameter set NL3 for neutron rich even-even
Ni isotopes. Results show that the contribution of the proper
treatment of the resonant continuum to pairing correlations for
those nuclei close to neutron drip line is important. The pairing
gaps, Fermi energies, pairing correlation energies, and binding
energies are considerably affected with a proper consideration of
the width of resonant states. The problem of unphysical particle
gas, which may appear in the calculation of the traditional mean
field plus BCS method for nuclei in the vicinity of drip line
could be well overcome when the pairing correlation is performed
by using the resonant states instead of the discretized states in
the continuum.
\end{abstract}


\maketitle

\section{introduction}
The study of properties of exotic nuclei, in particular the
structure of nuclei far from the $\beta$-stability line, has
recently attracted wide interest both experimentally and
theoretically. In those nuclei, the presence of nucleons, which
have separation energies appreciably smaller than those in ${\beta
}$-stable nuclei and have a far extending nucleon distribution,
leads to various interesting physical phenomena. Due to the
closeness of the Fermi surface to the particle continuum in exotic
nuclei the coupling between bound states and the continuum become
important. The description of exotic nuclei has to explicitly
include the coupling between bound states and the continuum. There
exist several extensive studies of exotic nuclei in relativistic
and non-relativistic microscopic
approaches\cite{Ring96,Dob98,Kuo97,Dob96}. A key ingredient of
those models is how to properly treat the pairing correlations
which have an important influence on physical properties in exotic
nuclei. In general, pairing correlations in open shell nuclei can
be treated by the Bardeen-Cooper-Schrieffer (BCS) theory or
through the Bogoliubov transformation\cite{Ring80}. The main
feature of the Bogoliubov transformation, compared with the simple
BCS theory, is that the Hartree-Fock equation  or Dirac equation
and the gap equations are solved simultaneously with
self-consistent fields. Both of them could give a good description
of the pairing correlation if the nucleus is not too far away from
the $\beta$-stable line\cite{Li02}. The simple BCS method may not
be reliable near the drip line because the continuous states were
not correctly treated\cite{Li02,Dob84}. The contribution of the
coupling to the continuum would be prominent when the nucleus
close to the drip line, therefore a proper treatment of the
continuum becomes more important. Single particle resonant states
in the continuum, which differ from bound states, are metastable
states with life times due to a sufficiently high centrifugal
barrier (for neutrons and protons) and Coulomb barrier (for
protons).  The effect of the widths of resonant states on pairing
correlations was not estimated by the usual non-relativistic
Hartree-Fock-Bogoliubov (HFB)\cite{Dob96,Ring80} or the
relativistic Hartree-Bogoliubov (RHB)\cite{Ring91}. The effect of
a proper treatment of the resonant continuum on the pairing in the
non-relativistic approach has been
investigated\cite{San97,San00,Kru01}. The investigation in the
relativistic approach is reported very recently\cite{San032}.

In this paper, we aim at the investigation on the effect of the
resonant continuum on pairing correlations of neutron rich nuclei
in the relativistic mean field theory plus BCS (RMF+BCS)
approximation. Particularly we shall focus our attention on the
effect of widths of single particle resonant states in the
description of neutron rich nuclei. It has been pointed out that
pairing correlations could be well described by the simple BCS
theory if single particle states in the continuum are properly
treated\cite{San97,San00,Kru01}. It is known that narrow resonant
states in the continuum are well localized inside the nucleus due
to a sufficiently high centrifugal barrier and Coulomb barrier
(for proton only). In our previous work\cite{Cao02}, we have shown
that the nuclear dynamical processes can also be well described by
the simple method where the continuum was replaced by only a few
single particle resonant states. Therefore, we may expect that
using a few single particle resonant states instead of the
discretized continuum, the pairing space would be large enough in
order to give a good description on the ground state properties of
neutron rich nuclei in the RMF, even within the BCS approximation.

Usually pairing correlations in the BCS approximation are treated
in a phenomenological way based on empirical pairing gaps deduced
from odd-even mass differences. This procedure works well in the
valley of $\beta$-stability, where experimental masses are known.
The predictive power for pairing gaps for nuclei far from the
$\beta$-stability  line is thus limited due to the unknown masses
of such nuclei. Therefore, a constant pairing strength is usually
employed in the BCS calculation\cite{Est01}.  We shall investigate
the pairing effect in the RMF with the parameter set
NL3\cite{Lal97}, which give a good description of not only the
ground state properties\cite{Las99} but also the collective giant
resonance\cite{Ma01,Ma02,Ring01,Cao03}. To investigate the width
effect of single particle resonant states  on pairing
correlations, we compare the results in three types of
calculations: (a) the RMF+BCS approach where quasi-particle wave
functions are obtained by a box boundary condition; (b) the
extended RMF+BCS approximation where wave functions of resonant
states are calculated by imposing a proper scattering boundary
condition for the continuous spectrum but the width of resonant
states is not considered, which is denoted as RMF+BCSR; (c) the
full RMF+BCS (RMF+BCSRW) calculation where widths of single
particle resonant states are taken into account explicitly.

The paper is arranged as the follows. In Sec.II and Sec.III the
relativistic mean field theory and the BCS approximation with
explicitly considering the widths of the resonant continuum are
presented. The effect of the resonant continuum on pairing
correlations is  formulated by introducing a continuum level
density into the gap equations. The results calculated by
different models mentioned above for neutron rich even-even Ni
isotopes are given and discussed in Sec.IV. Finally we give a
brief summary in Sec.V.

\section{Relativistic Mean Field Theory}

 The RMF theory \cite{Ring96,Hor81,Ser86}
is based on the effective Lagrangian density which  includes the
nucleon field ($\psi$), the isoscalar scalar meson field
($\sigma$), the isoscalar vector meson field ($\omega$), the
isovector vector meson field ($\bff{\rho}$), and electromagnetic
field ($A$). The Lagrangian density is expressed as the following:
\begin{eqnarray}
\cal{L} &=&\overline{\psi }(i\gamma^{\mu} \partial_{\mu} -M)\psi
+\frac{1}{2}\partial _\mu \sigma \partial ^\mu \sigma -U(\sigma
)-\frac{1}{4}\Omega _{\mu \nu }\Omega ^{\mu \nu
} \nonumber \\
&&+\frac{1}{2}m_\omega^2 \omega_{\mu}\omega^{\mu}- \frac{1}{4}
{\bff{R}}_{\mu \nu }{\bff{R}}^{\mu \nu }+\frac 12m_\rho ^2
{\bff{\rho}}_{\mu}{\bff{\rho}}^{\mu}-\frac 14F_{\mu \nu }F^{\mu
\nu } \nonumber \\
&&-g_\sigma \overline{\psi}\sigma \psi -g_\omega \overline{\psi
}\gamma^{\mu} \omega_{\mu} \psi -g_\rho \overline{\psi
}\gamma^{\mu} {\bff{\rho}}_{\mu} \cdot {\bff{\tau}}\psi \nonumber
\\  &&-e\overline{\psi }\gamma^{\mu} A_{\mu}\left(
\frac{1-\tau_3}2\right) \psi ~,   \label{eq1}
\end{eqnarray}
where $M$, $m_\sigma$, $m_\omega$, and $m_\rho$ denote the
nucleon, $\sigma$, $\omega$, and $\rho$ mesons masses,
respectively, while $g_\sigma$, $g_\omega$ and $g_\rho$ are the
corresponding coupling constants for the mesons, respectively. The
vectors in isospin space are denoted by bold-faced symbols.
$\Omega ^{\mu \nu }$, ${\bff{R}}^{\mu \nu}$ and $F^{\mu \nu }$ are
the field tensors of the vector fields $\omega $, $\rho $ and of
the photon, respectively, they are defined as:
\begin{equation}
\Omega^{\mu\nu}=\partial^{\mu}\omega^{\nu}-\partial^{\nu}\omega^{\mu}
~, \label{eq2}
\end{equation}
\begin{equation}
{\bff{R}}^{\mu\nu}=\partial^{\mu}{\bff{\rho}}^{\nu}-\partial^{\nu}{\bff{\rho}}^{\mu}
~, \label{eq3}
\end{equation}
\begin{equation}
F^{\mu\nu}=\partial^{\mu}A^{\nu}-\partial^{\nu}A^{\mu} ~,
\label{eq4}
\end{equation}

In the Eq.(1), a non-linear scalar self-interaction
term\cite{Bog77} $U(\sigma )$ of the $\sigma $ meson has been
taken into account:

\begin{equation}
U(\sigma )=\frac 12m_\sigma ^2\sigma ^2+\frac 13g_2\sigma ^3+\frac
14 g_3\sigma ^4 ~,\label{eq5}
\end{equation}

 The Dirac equation of a single-particle
state in the RMF approximation can be expressed as following:
\begin{equation}
\left[ {\bff{\alpha}} \cdot {\bff{p}}+V(r)+\beta \left(
M-S(r)\right) \right] \psi _{\alpha}=E_{\alpha}\psi _{\alpha}~,
 \label{eq6}
\end{equation}
where  $E_{\alpha}$ and $\psi _{\alpha}$ are the single-particle
energy and wave function, respectively. $S(r)$ and $V(r)$ are an
attractive scalar and repulsive vector potential, respectively. In
the RMF approximation, scalar and vector potentials are produced
by the classical meson fields: isoscalar $\sigma$, $\omega$ and
isovector $\rho$ mesons as well as the photon, which are obtained
in a self-consistent calculation for the nuclear ground
state\cite{Ring96,Hor81,Ser86}. The nucleon spinor $\psi_{\alpha}$
in the Dirac equation is expressed as:
\begin{equation}
\psi _\alpha ({\bff{r}})=\frac{1}{r} \left( \begin{array}{c}
  iG_a(r) \\
  F_a(r){\bff{\sigma}}\cdot {\bff{\hat{r}}}
\end{array} \right)
\Phi_{\kappa,m} ( \hat{r})\chi _{\frac 12}~, \label{eq7}
\end{equation}
where $G_{a}$ and $F_{a}$ are upper and lower components of
nucleon wave function, $\chi _{\frac 12}$ is the isospinor and
$\Phi_{\kappa,m}$ is a spherical harmonics function. $\alpha $ is
a set of quantum numbers $\alpha =(n,l,j,m,\tau
_3)\equiv(a,m,\tau_3)$. $\kappa$ is the Dirac quantum number given
by:
\begin{eqnarray}
 \kappa =
  \left\{
    \begin{array}{lll}
      -(j+\frac{1}{2}),
     & \text{for}\ & j=l+\frac{1}{2} \\
         +(j+\frac{1}{2}),
     & \text{for}\   &j=l-\frac{1}{2} \\

    \end{array}
  \right.
 ~. \label{eq8}
\end{eqnarray}

For spherical nuclei, the Dirac equation can be reduced to coupled
equations of the radial part for $G_a(r)$ and $F_a(r)$:
\begin{widetext}
\begin{equation}
\frac{d}{dr}
  \left( \begin{array}{c}
    G_a(r) \\
    F_a(r) \
  \end{array} \right) =\left(\begin{array}{cc}
    -\frac{\kappa}{r} & M+E_a-S(r)-V(r)) \\
    M-E_a-S(r)+V(r) & \frac{\kappa}{r} \
  \end{array}\right)\left(\begin{array}{c}
    G_a(r) \\
    F_a(r) \
  \end{array}\right)~,
\label{eq9}
\end{equation}
\end{widetext}
where $S(r)$ is the scalar potential:
\begin{equation}
S(r)=g_{\sigma}\sigma(r) ~, \label{eq10}
\end{equation}
and $V(r)$ denotes the vector potential:
\begin{equation}
V(r)=g_{\omega}\omega_{0}(r)+g_{\rho}\tau_{3}\rho_{0}(r)+e\frac{1-\tau_{3}}{2}A_{0}(r)
~, \label{eq11}
\end{equation}

 The meson and electromagnetic fields obey the radial
Laplace equations:
\begin{equation}
\frac{d^{2}}{dr^{2}}\sigma(r)+\frac{2}{r}\frac{d}{dr}\sigma(r)
-m_{\sigma}^{2}\sigma(r)=-g_{\sigma}\rho_{s}(r)-g_{2}\sigma^{2}(r)-g_{3}\sigma^{3}(r)
~, \label{eq12}
\end{equation}
\begin{equation}
\frac{d^{2}}{dr^{2}}\omega_{0}(r)+\frac{2}{r}\frac{d}{dr}
\omega_{0}(r)-m_{\omega}^{2}\omega_{0}(r)=-g_{\omega}\rho_{v}(r)
~, \label{eq13}
\end{equation}
\begin{equation}
\frac{d^{2}}{dr^{2}}\rho_{0}(r)+\frac{2}{r}\frac{d}{dr}\rho_{0}(r)
-m_{\rho}^{2}\rho_{0}(r)=-g_{\rho}\rho_{3}(r)~, \label{eq14}
\end{equation}
\begin{equation}
\frac{d^{2}}{dr^{2}}A_{0}(r)+\frac{2}{r}\frac{d}{dr}A_{0}(r)=-e\rho_{c}(r)
~, \label{eq15}
\end{equation}
with
\begin{equation}
 \rho_{s}(r) =  \sum_{a}^{occ}\frac{2j_{a}+1}{4\pi}(|G_a(r)|^2 - |F_a(r)|^2) ~,
\label{eq16}
 \end{equation}
\begin{equation}
 \rho_v(r) =  \sum_{a}^{occ}\frac{2j_{a}+1}{4\pi}(|G_a(r)|^2 + |F_a(r)|^2) ~,
\label{eq17}
 \end{equation}
\begin{equation}
 \rho_3(r)  =  \sum_{a}^{occ}\frac{2j_{a}+1}{4\pi}\tau_{3}(|G_a(r)|^2 + |F_a(r)|^2) ~,
 \label{eq18}
 \end{equation}
\begin{equation}
 \rho_c(r)  =  \sum_{a}^{occ}\frac{2j_{a}+1}{4\pi}\left(\frac{1-\tau_{3}}{2}\right)
 (|G_a(r)|^2 +|F_a(r)|^2) ~,
 \label{eq19}
\end{equation}

The sums $a$ run over all occupied states. The Dirac equation
Eq.(9) and the corresponding meson field functions Eq.(12 - 14) as
well as the photon field function Eq.(15) with expressions for
source terms (16 - 19) can be solved self-consistently. More
details on the relativistic mean-field theory can be found in
Refs.\cite{Ring96,Hor81,Ser86}.

\section{BCS Approximation with Resonant Continuum}

It is well known that the pairing correlation plays an important
role in describing the ground state properties of open shell
nuclei. The non-relativistic HFB\cite{Dob96,Ring80} or
RHB\cite{Ring91} theory can provide a unified description on the
mean field and pairing correlations, but they require much more
numerical effort than the simple BCS theory. However, the width
effect of resonant states in the gap equation is missing in most
previous works of the BCS theory or the Bogoliubov transformation.
Recently, properly treating the contribution of resonant continuum
to pairing correlations has been investigated in the
non-relativistic HF+BCS as well as HFB in
Refs.\cite{San97,San00,Kru01,Gra01}. It has been pointed
out\cite{Gra01} that the resonant continuum HF+BCS approximation
with a proper treatment
of resonant continuum
can reproduce the pairing correlation energies predicted by the
continuum HFB approach up to the drip line.
In this paper, we investigate the width effect of single particle
resonant states on the pairing correlations in the relativistic
approach. The pairing correlations of open shell nuclei are
treated based on the RMF within the BCS approximation. A constant
pairing strength is adopted in the BCS theory. Although the
extended RMF+BCS theory formulated with correct boundary
conditions for the continuous spectrum is available in the
literature\cite{San032}, in the following we briefly summarize the
essential points together with our model. From Ref.\cite{Ring80},
assuming constant pairing matrix elements in the vicinity of Fermi
level, one can obtain the so-called gap equations:
\begin{equation}
\sum_{a}\left(j_{a}+\frac{1}{2}\right)\frac{1}{\sqrt{(\varepsilon_{a}-\lambda)^{2}+\Delta^{2}}}
= \frac{2}{G} ~,
\label{eq20}
\end{equation}
\begin{equation}
\sum_{a}\left(j_{a}+\frac{1}{2}\right)\left[1 -
\frac{\varepsilon_{a}-\lambda}{\sqrt{(\varepsilon_{a}-\lambda)^{2}
+\Delta^{2}}}\right] = N ~,
\label{eq21}
\end{equation}
where $\lambda$, $\Delta$, and $G$ represent the Fermi energy,
pairing gap, and the pairing force constant, respectively.  $N$ is
the number of neutrons or protons involved in the pairing
correlations. The solution of those two coupled equations, Eq.(20)
and (21) allows one to find $\lambda$ and $\Delta$. The
contribution of the pairing interaction to the total energy can be
expressed as following:
\begin{equation}
E_{pair} =-G\left[\sum_{a}\left(j_{a}+\frac{1}{2}\right)v_{a}u_{a}
\right]^{2} ~, \label{eq22}
\end{equation}
where $v_{a}^{2}$ is the occupation probability of a state with
quantum numbers $\alpha =(n,l,j,m,\tau _3)\equiv(a,m,\tau_3)$ and
$u_{a}^{2}=1-v_{a}^{2}$. For spherical nuclei, the eigenvalues of
Dirac equations are of degeneracy for magnetic quantum numbers $m$
within the same $(n,l,j)$. One has:
\begin{equation}
\left( \begin{array}{c}
    v_{a}^{2} \\
    u_{a}^{2} \
  \end{array}
  \right)=\frac{1}{2}\left(1\mp\frac{\varepsilon_{a}-\lambda}
  {\sqrt{(\varepsilon_{a}-\lambda)^{2}+\Delta^{2}}}
  \right) ~,
 \label{eq23}
\end{equation}

In principle, the continuous spectrum is also the solution of
Dirac Eq.(9), both continuous and bound states constitute a
complete set of basis. In most previous theoretical nuclear
structure calculations, single particle states in the continuum
are usually treated in a discretization procedure by expanding
wave functions in a set of harmonic oscillator basis or setting a
box. This approximation, however, can be justified for very narrow
resonances and gives a global description of the contributions
from the continuum. In this work we introduce single particle
resonant states into the pairing gap equations 
instead of the discretized continuous states in order to
investigate the width effect of resonances on pairing
correlations. The resonant state wave function are obtained by
imposing a scattering boundary condition. At the distance R where
the nuclear potentials vanish, the upper component of the neutron
radial wave function has the following asymptotic behavior:
\begin{equation}
G_\nu(kr)=A_\nu\left[ j_{l_{\nu}} (kr)-\tan \delta _\nu
n_{l_{\nu}}(kr)\right]~, ~~~~~ {\textrm{for}}~~  r\geq R~.
\label{eq24}
\end{equation}
where $j_{l_{\nu}}$ and $n_{l_{\nu}}$ are spherical Bessel and
Neumann functions, respectively, and $\delta_{\nu}$ is the phase
shift corresponding to the angular momentum $(l_{\nu},j_{\nu})$,
$k^{2}= E^{2}-M^{2} $. For the case of proton, the asymptotic
behavior can be obtained by replacing the spherical Bessel and
Neumann functions in Eq.(24) with the relativistic regular and
irregular Coulomb wave functions\cite{Gre90}, respectively. The
energy of a resonant state is determined when the phase shift of
the scattering state reaches $\pi /2$. The wave function of
scattering state is normalized to a delta function of energy
$\delta(E-E')$. This condition fixes the normalization constant
$A_\nu$, it can be obtained as:
\begin{equation}
A_{\nu}=\cos\delta_\nu\sqrt{\frac{1}{\pi}\frac{M+E}{2k}}
~,\label{eq25}
\end{equation}

In order to take into account the width effect of resonant states
in the continuum we introduce a level density of the
continuum\cite{Bon84} into the gap equations. When we introduce
single particle resonant states with widths into the pairing
correlations, the gap equations can be expressed as:
\begin{widetext}
\begin{equation}
\sum_{a}\left(j_{a}+\frac{1}{2}\right)\frac{1}{\sqrt{(\varepsilon_{a}-\lambda)^{2}+\Delta^{2}}}
+ \sum_{\nu} \left(j_{\nu}+\frac{1}{2}\right)
\int_{I_{\nu}}\frac{g_{\nu}(\varepsilon_{\nu})}
{\sqrt{(\varepsilon_{\nu}-\lambda)^{2}+\Delta^{2}}}d\varepsilon_{\nu}
 = \frac{2}{G} ~,
\label{eq26}
\end{equation}

\begin{equation}
\sum_{a}\left(j_{a}+\frac{1}{2}\right)\left[1 -
\frac{\varepsilon_{a}-\lambda}{\sqrt{(\varepsilon_{a}-\lambda)^{2}+\Delta^{2}}}\right
] + \sum_{\nu} \left(j_{\nu}+\frac{1}{2}\right)
\int_{I_{\nu}}g_{\nu}(\varepsilon_{\nu})\left[1 -
\frac{\varepsilon_{\nu}-\lambda}{\sqrt{(\varepsilon_{\nu}-\lambda)^{2}+\Delta^{2}}}\right
]d\varepsilon_{\nu} = N ~.
\label{eq27}
\end{equation}
\end{widetext}
The sums $a$ and $\nu$ run over the bound states and resonant
states involved in the pairing  calculation, respectively, and
$I_{\nu}$ is an energy interval associated with each partial wave
$(l_{\nu},j_{\nu})$. The factor $g_{\nu}$ is defined as:
\begin{equation}
g_{\nu}(\varepsilon_{\nu}) =
\frac{1}{\pi}\frac{d\delta_{\nu}}{d\varepsilon_{\nu}}
~,\label{eq28}
\end{equation}
which is the so-called continuum level density and $\delta_{\nu}$
is the phase shift of the scattering state with angular momentum
$\nu = (l_{\nu},j_{\nu})$. The factor $g_{\nu}$ represents the
variation of the localization of scattering states in the energy
region of a resonance, in other words, it reflects the widths
effect of the resonant states. For a very narrow resonant state,
the factor $g_{\nu}$ becomes a delta function.

Taking account of those resonant continuum in the gap
equations, the expressions of various densities in Eq.(16 - 19)
have to be modified. They can be expressed as:
\begin{widetext}
\begin{eqnarray}
\rho_s(r)&=&\sum_{a}\frac{2j_{a}+1}{4\pi}v_{a}^{2}(|G_{a}(r)|^2-|F_{a}(r)|^2)
\nonumber \\
&&+\sum_{\nu}\frac{2j_{\nu}+1}{4\pi}\int_{I_{\nu}}(|G_\nu(r)|^2-|F_\nu(r)|^2)
g_{\nu}(\varepsilon_{\nu})v_{\nu}^{2}(\varepsilon_{\nu})d\varepsilon_{\nu}
~,\label{eq29}
\end{eqnarray}
\begin{eqnarray}
\rho_v(r)&=&\sum_{a}\frac{2j_{a}+1}{4\pi}v_{a}^{2}(|G_{a}(r)|^2+|F_{a}(r)|^2)
\nonumber \\
&&+\sum_{\nu}\frac{2j_{\nu}+1}{4\pi}\int_{I_{\nu}}(|G_\nu(r)|^2+|F_\nu(r)|^2)
g_{\nu}(\varepsilon_{\nu})v_{\nu}^{2}(\varepsilon_{\nu})d\varepsilon_{\nu}
~,\label{eq30}
\end{eqnarray}
\begin{eqnarray}
\rho_3(r)&=&\sum_{a}\frac{2j_{a}+1}{4\pi}\tau_{3}v_{a}^{2}(|G_{a}(r)|^2+|F_{a}(r)|^2)
\nonumber \\
&&+\sum_{\nu}\frac{2j_{\nu}+1}{4\pi}\tau_{3}\int_{I_{\nu}}(|G_\nu(r)|^2+|F_\nu(r)|^2)
g_{\nu}(\varepsilon_{\nu})v_{\nu}^{2}(\varepsilon_{\nu})d\varepsilon_{\nu}
~,\label{eq31}
\end{eqnarray}
\begin{eqnarray}
\rho_c(r)&=&\sum_{a}\frac{2j_{a}+1}{4\pi}\left(\frac{1-\tau_{3}}{2}\right)v_{a}^{2}
(|G_{a}(r)|^2+|F_{a}(r)|^2) \nonumber \\
&&+\sum_{\nu}\frac{2j_{\nu}+1}{4\pi}\left(\frac{1-\tau_{3}}{2}
\right)\int_{I_{\nu}}(|G_\nu(r)|^2+|F_\nu(r)|^2)g_{\nu}(\varepsilon_{\nu})v_{\nu}^{2}(\varepsilon_{\nu})d\varepsilon_{\nu}
~,\label{eq32}
\end{eqnarray}

The Dirac equation Eq.(9) and  meson field functions (12 - 14) as
well as the photon field function (15) with corresponding
densities (29 - 32) are solved self-consistently in an iterative
way. Therefore, the total binding energy can be written as:
\begin{eqnarray}
E&=&\sum_{a}\left(2j_{a}+1\right)v_{a}^{2}E_{a}+\sum_{\nu}\left(2j_{\nu}+1\right)
\int_{I_{\nu}}g_{\nu}(\varepsilon_{\nu})v_{\nu}^{2}(\varepsilon_{\nu})
\varepsilon_{\nu}d\varepsilon_{\nu} \nonumber \\
&&-\frac{1}{2}\int\left(g_{\sigma}\rho_{s}\sigma_{0}+g_{\omega}\rho_{\nu}\omega_{0}+g_{\rho}\rho_{3}\rho^{3}_{0}+e\rho_{c}A_{0}
\right)d^{3}r \nonumber \\
&&-\int\left(\frac{1}{3}g_{2}\sigma_{0}^{3}+\frac{1}{4}g_{3}\sigma_{0}^{4}\right)d^{3}r \nonumber \\
&&-G\left[\sum_{a}\left(j_{a}+\frac{1}{2}\right)v_{a}u_{a}+\sum_{\nu}\left(j_{\nu}+\frac{1}{2}\right)\int_{I_{\nu}}g_{\nu}(\varepsilon_{\nu})v_{\nu}(\varepsilon_{\nu})u_{\nu}(\varepsilon_{\nu})d\varepsilon_{\nu}
\right]^{2} \nonumber \\
&&-\frac{3}{4}\cdot41\cdot A^{-1/3}
 ~,\label{eq33}
\end{eqnarray}
Where the last two terms are the pairing energy and the correction
for the spurious center of mass motion, respectively.
\end{widetext}

\section{Properties of neutron rich even-even Ni isotopes}

In this work we investigate the ground state properties of neutron
rich even-even Ni isotopes in the RMF+BCS model and explicitly
take account of the continuum. Particularly we shall focus our
attention on the effect of single particle resonant states on the
pairing correlation in those neutron rich nuclei. Calculations are
carried out in the RMF  with the parameter set NL3.  Three types
of calculations: RMF+BCS, RMF+BCSR and RMF+BCSRW are performed.
The effect of single particle states in the continuum on the
ground state properties is investigated.

In our investigation, the proton number  of Ni isotopes is $Z=28$,
which is a closed shell. Therefore, the proton pairing gap is
taken to be zero. For the neutron case, we use a state-independent
pairing strength $G=C/A$, where the constant $C=20.5$ MeV is
adopted from Ref.\cite{Est01}. It could also provide a best fit to
reproduce experimental binding energies of known Ni-isotopes. In
practical calculations, we restrict the pairing space to about one
harmonic oscillator shell above and below the Fermi surface in the
RMF+BCS model. The levels include $1f_{5/2}$, $2p_{3/2}$,
$2p_{1/2}$,$1g_{9/2}$, $2d_{5/2}$, $3s_{1/2}$, $2d_{3/2}$,
$1g_{7/2}$, $1h_{11/2}$ and $2f_{7/2}$,
some of them, such as $2d_{5/2},~ 2d_{3/2},~ 1g_{7/2}, ~
2f_{7/2}$, and $1h_{11/2}$ are single particle resonant states,
depending on the isotopes. Highly excited resonant states with
large widths, such as $1h_{9/2}$ and $1i_{13/2}$, are ignored in
our calculations. Obviously, an enlarged pairing space in the
normal BCS calculation would change the pairing contribution due
to a constant pairing interaction. Therefore the results in cases
of RMF+BCS and RMF+BCSR in Fig.1 and Fig.2 would be different if
one includes more states in the pairing active space. It is found
that the pairing properties are not affected too much if one
includes the resonant states $1h_{9/2}$ and $1i_{13/2}$ with wide
widths in the RMF+BCSRW calculation.

The single particle energies and wave functions are first carried
out by solving the Dirac equation Eq.(9) self-consistently. Using
those single particle states we can solve the BCS gap equations in
the three approaches. The Fermi energy and gap as well as the
occupation probabilities of quasi-particle states are obtained
simultaneously. The nuclear densities composed of quasi-particle
states and potentials are
recalculated. Then we solve Dirac equation again by an iterative
way until the convergence is reached. The pairing correlation
energies can be obtained in the Eq.(22) in the three approaches,
which are shown in Fig.1 for neutron rich even-even Ni isotopes
of $N=40$ to 70. Pronounced differences of the pairing 
energies performed in various RMF+BCS approaches are observed for
open shell nuclei, where the behavior of pairing 
energies is similar to those obtained by the RHB in
Refs.\cite{Meng98}. The usual RMF+BCS approach produces the
largest pairing energies, 
due to some non-resonant
scattering states in the continuum are included. It can be seen in
Fig.1 that the pairing 
energies are reduced largely
when the widths of single particle resonant states are taken into
account in  the RMF+BCSRW calculation.

In order to understand the difference presented by different
treatments on pairing correlations, we plot the results of pairing
gaps and Fermi energies for neutron rich even-even Ni isotopes
with masses from 68 to 98 in Fig.2(a) and Fig.2(b). The curve in
Fig.2 (a) is the empirical pairing gap given by $11.2/\sqrt{A}$.
It is observed that pairing gaps are largely decreased when  the
width effect of single particle resonant states is taken in
account in pairing correlation calculations, while the Fermi
energies remain unchanged for three approaches. This feature
agrees with the case in the non-relativistic HF+BCS
calculations\cite{San00}. It is found that the pairing gaps in the
RMF+BCSR calculation are close to those obtained in the box
RMF+BCS approximation. As an example, for the neutron rich nucleus
$^{90}$Ni, the pairing gap and Fermi energy are $\Delta_{n}$
=1.558 MeV and $\lambda_{n}$=-1.278 MeV in the RMF+BCSR approach,
whereas, they are reduced to 1.223 MeV and -1.256 MeV,
respectively when we perform the pairing correlation calculations
by considering the width effect of resonances explicitly. The
reduced pairing gap and Fermi energy could change the occupation
probability of neutrons at each single particle orbit near the
Fermi surface. In Fig.3 we show the occupation probabilities of
neutron single particle levels in $^{90}$Ni as a function of
corresponding energies. The arrow is the position of the neutron
Fermi surface. In $^{90}$Ni the Fermi energy is very close to the
continuum, therefore the continuum becomes important in the
pairing correlations. Some scattering states in the continuum,
such as states 3$p_{3/2}$ and 3$p_{1/2}$, which are not resonance
states due to the low centrifugal barrier, are included in the
RMF+BCS calculation. Therefore the RMF+BCS approach overestimates
the pairing correlation and produces large pairing energies and
pairing gaps. It is found that the width effect on the pairing is
to reduce the pairing correlations, therefore occupation
probabilities of low-lying states below the Fermi surface in the
case of RMF+BCSRW
are slightly larger
than those without the width effect, and vice versa for states
above Fermi surface.

In Fig.4 we display the two-neutron separation energy $S_{2n}$:
\begin{equation}
S_{2n}=B(Z,N)-B(Z,N-2) ~. \label{eq34}
\end{equation}
We calculate two-neutron separation energies for Ni isotopes till
the neutron drip line in three approaches, RMF+BCS, RMF+BCSR and
RMF+BCSRW. The two-neutron separation energies calculated in the
RMF without pairings are also plotted in Fig.4, which are denoted
by crosses. The empirical data of $S_{2n}$ are obtained by
Eq.(34), where the experimental binding energies of the Ni
isotopes for $N\leq$50 are taken from the Ref.\cite{Audi95}. The
theoretical binding energies for Ni isotopes are listed in Table
I. The position of the neutron drip line may be determined by the
signature where the two-neutron separation energy changes its sign
or the Fermi energy becomes positive. It is found that the
position of the neutron drip line predicted by various RMF+BCS
models is located between $^{98}$Ni and $^{100}$Ni, the same
result is also obtained in RHB calculations\cite{Meng98}. It is
found that the two-neutron separation energies calculated in
various RMF+BCS approaches are very close to each other, which can
well reproduce the experimental data in the region of $N=42$ to
$N=50$. In contrast, the results without pairing considerably
deviate from the others. It indicates that the pairing is
responsible to a reasonable description of the two-neutron
separation energy. Some disagreements with experimental data are
found for $N=38$ and $N=40$ isotopes, which were also observed in
the RHB calculation with the parameter set NL3\cite{Lal98,Sha00}.
Although pairing energies obtained in these three approaches with
different treatment of pairings differ significantly, two-neutron
separation energies are consistent with each other. This is
because that differences, which appear in the pairing correlation
energy are largely cancelled in the $S_{2n}$.

In table I, it is shown that the binding energies produced in the
three RMF+BCS approaches scatter when nuclei are far away from the
$\beta$ stable line. To clearly illustrate the width effect of the
resonant states on the pairings we define a new quantity, which
characterizes the pairing correlations energy:
\begin{equation}
E_{BCS}=E(RMF)-E(RMF+BCS) ~. \label{eq35}
\end{equation}
The $E_{BCS}$ of Ni isotopes calculated in various approaches are
plotted in Fig.5. It is shown that the results produced in the
three approaches are very close to each other for those isotopes
not far away from the $\beta$ stable-line. It implies that the
width of the resonance in the pairing correlation could be
neglected at $N\leq 34$. The RMF+BCS and RMF+BCSR give more or
less similar values of $E_{BCS}$, even near the drip line. The
width effect gets more and more pronounced as the neutron number
increases, especially near the drip line. Therefore a proper
treatment of the resonant continuum including its width might be
necessary for the nuclei near the drip line.

Neutron rms radii of neutron rich even-even Ni isotopes are
further calculated in the three approaches, which are plotted in
Fig.6. The rms radii calculated in the RMF are also shown in Fig.6
and displayed by cross. It is seen that neutron rms radii of Ni
isotopes at $A<80$ calculated in all approaches are very close to
each other, which is similar to what was observed in the binding
energy. The neutron rms radii calculated in the RMF+BCS with a box
become larger at $A>80$ in comparison with those produced in other
approaches. This is due to the fact that the contribution of the
pairing correlation from the scattering states $3p_{3/2}$ and
$3p_{1/2}$ is included in the RMF+BCS calculations. They are not
resonant states and their wave functions are not well localized
inside the nucleus. The neutron rms radius for $^{84}$Ni given in
the RMF is slightly smaller than that obtained in the other
approaches. Actually this is a natural result when the pairing
correlation is switched on. The neutron rms radius in the RMF+BCS
is increased because the pairing interaction scatters some
neutrons from $2d_{5/2}$ to the loosely bound state $3s_{1/2}$
which is a state less localized. In contrast, in the case of
$^{90}$Ni the states $3s_{1/2}$ and $2d_{3/2}$, which are
completely occupied in the RMF approximation are scattered to
states $1g_{7/2}$ and $1h_{11/2}$ due to the pairing interaction.
Although the states $1g_{7/2}$ and $1h_{11/2}$ are close to or
buried in the continuum, their wave functions are more localized
inside the nucleus than those for the $3s_{1/2}$ and $2d_{3/2}$
states due to a high centrifugal barrier. Therefore the rms radius
of $^{90}$Ni calculated in the RMF is larger than that produced in
calculations including the pairing correlation.

We plot the neutron density of $^{84}$Ni in Fig.7, where the
dash-dotted curve is obtained in the RMF, the short-dotted, dashed
and solid curves are produced in the RMF+BCS, RMF+BCSR and
RMF+BCSRW approaches, respectively.  The tail of the density
distribution gets larger when the pairing correlation is
considered. The width effect on the density distribution is very
small, which is consistent to the calculated neutron rms radii. In
order to show the effect of the box size on the density in the box
RMF+BCS approximation we also plot the neutron density with
various values of the box size R$_{\text{box}}$ = 15, 20, 25 fm in
Fig.7. It is observed that the tail of the density strongly
depends on the box size, where unphysical particle gas may appear
in the exotic nucleus. In the RMF+BCS calculation non-resonant
discretized states in the continuum are included in pairing
correlation, such as $3p_{3/2}$ and $3p_{1/2}$ states. Their wave
functions have a long tail strongly depending on the box size. It
is found that this problem is well overcome when one performs the
pairing correlation calculation  with only a few narrow resonant
states in stead of those discretized states in the continuum.

\section{Summary}

In this paper, we have investigated the pairing correlation for
neutron-rich Ni isotopes in the relativistic mean field theory and
Bardeen-Cooper-Schrieffer approximation with a constant pairing
strength. A proper treatment of the single particle resonant state
in the continuum on pairing correlations has to include not only
its energy, but also its width. The inclusion of the width of the
resonant state in the pairings can be performed by introducing a
level density in the continuum into the pairing gap equation. The
resonant continuum is solved by imposing a proper scattering
boundary condition.  The investigation is performed in three
approaches: RMF+BCS, RMF+BCSR and RMF+BCSRW with the effective
Lagrangian parameter set NL3. A special attention is paid on the
width effect of resonant states in the continuum  on the pairing
correlation for nuclei close to the drip line. We have studied the
width effect of the resonant continuum on pairings: pairing gap,
Fermi energy and occupation probability, as well as nuclear ground
state properties, such as binding energy, two-neutron separation
energy, neutron rms radius and neutron density. The results show
that a proper treatment of the resonant continuum in pairing
correlations, which explicitly includes the width of the single
particle resonant state is important for nuclei close to neutron
drip line. They could affect the pairing gaps, Fermi energies,
pairing energies, and binding energies considerably. Various
RMF+BCS approaches could give a similar description on the ground
state properties for nuclei not far away from the $\beta$ stable
line. It is observed that unphysical particle gas, which may
appear in the traditional mean field plus BCS calculation for
nuclei in the vicinity of drip line can be well overcome when one
performs pairing correlation calculations with only resonant
states instead of discretized states in the continuum. It might be
concluded that the simple BCS method works well in describing the
pairing correlations for neutron rich nuclei provided the
continuum is properly treated.

\begin{acknowledgments}
One of the authors (L.G. Cao) wishes to thank Prof. Zhang Zong-ye
and Prof. Yu You-wen for many stimulating discussions. This work
is supported by the National Natural Science Foundation of China
under Grant Nos 10305014, 90103020, 10075080 and 10275094, and
Major State Basic Research Development Programme in China under
Contract No G2000077400.
\end{acknowledgments}

\newpage

\begin{table}
\caption{The binding energies of Ni isotopes near the neutron drip
line calculated in the RMF+BCS, RMF+BCSR, and RMF+BCSRW approaches
with parameter set NL3. The available experimental binding
energies of the Ni isotopes for $N\leq$50 from Ref.\cite{Audi95}
are also shown. All energy values in the table are in unit of
MeV.}
\begin{ruledtabular}
\begin{tabular}{ccccc}
 & RMF+BCS&RMF+BCSR&RMF+BCSRW&Exp.\\
\hline

$^{58}$Ni&502.347&502.275&502.104&506.453\\
$^{60}$Ni&521.655&521.448&521.363&526.841\\
$^{62}$Ni&540.458&540.374&540.251&545.258\\
$^{64}$Ni&558.361&558.273&558.055&561.754\\
$^{66}$Ni&575.642&575.579&575.009&576.830\\
$^{68}$Ni&590.892&590.960&590.930&590.430\\
$^{70}$Ni&603.233&603.195&602.775&602.600\\
$^{72}$Ni&614.246&614.116&613.567&613.900\\
$^{74}$Ni&624.428&624.307&623.752&623.900\\
$^{76}$Ni&633.840&633.816&633.414&633.100\\
$^{78}$Ni&642.352&642.555&642.555&641.400\\
$^{80}$Ni&647.837&647.659&646.964&\\
$^{82}$Ni&652.279&651.943&650.888&\\
$^{84}$Ni&656.092&655.678&654.472&\\
$^{86}$Ni&659.422&658.975&657.682&\\
$^{88}$Ni&662.284&661.863&660.582&\\
$^{90}$Ni&664.756&664.319&663.047&\\
$^{92}$Ni&667.756&666.456&665.254&\\
$^{94}$Ni&668.562&668.319&667.144&\\
$^{96}$Ni&670.133&670.183&669.332&\\
$^{98}$Ni&671.382&671.605&671.605&\\
\end{tabular}
\end{ruledtabular}
\end{table}

\begin{figure}[tbp]
\includegraphics[scale=0.5,angle=0.]{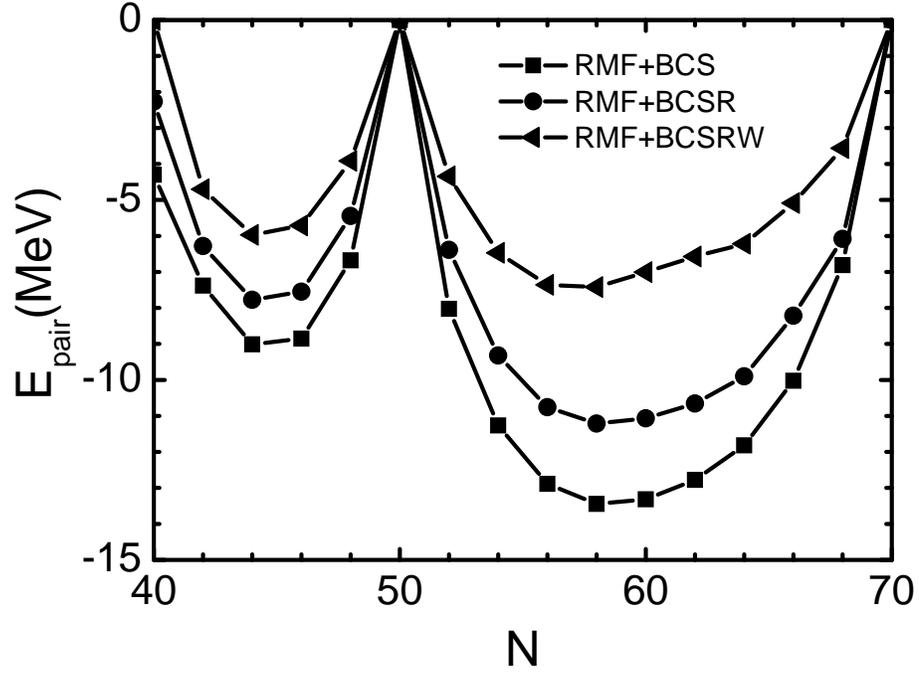}
\vglue -4.0cm \caption{The pairing energies for neutron rich
even-even Ni isotopes with $N=40-70$. The solid squares, circles
and left-triangles denote the results calculated in the RMF+BCS,
RMF+BCSR and RMF+BCSRW approaches, respectively. All calculations
are performed with the parameter set NL3. } \label{Fig1}
\end{figure}

\begin{figure}[tbp]
\includegraphics[scale=0.5,angle=0.]{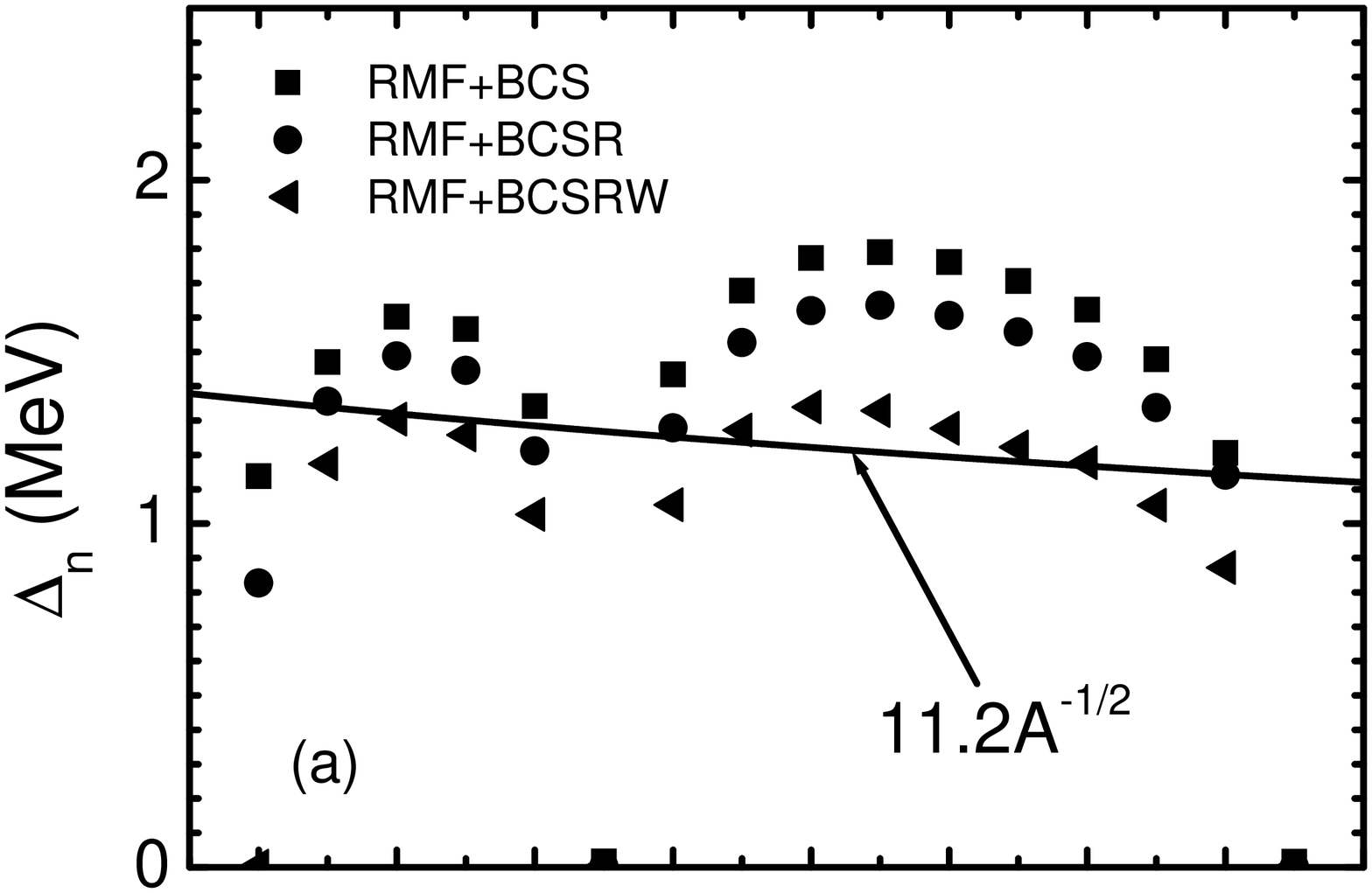}
\vglue -6.0cm
\includegraphics[scale=0.5,angle=0.]{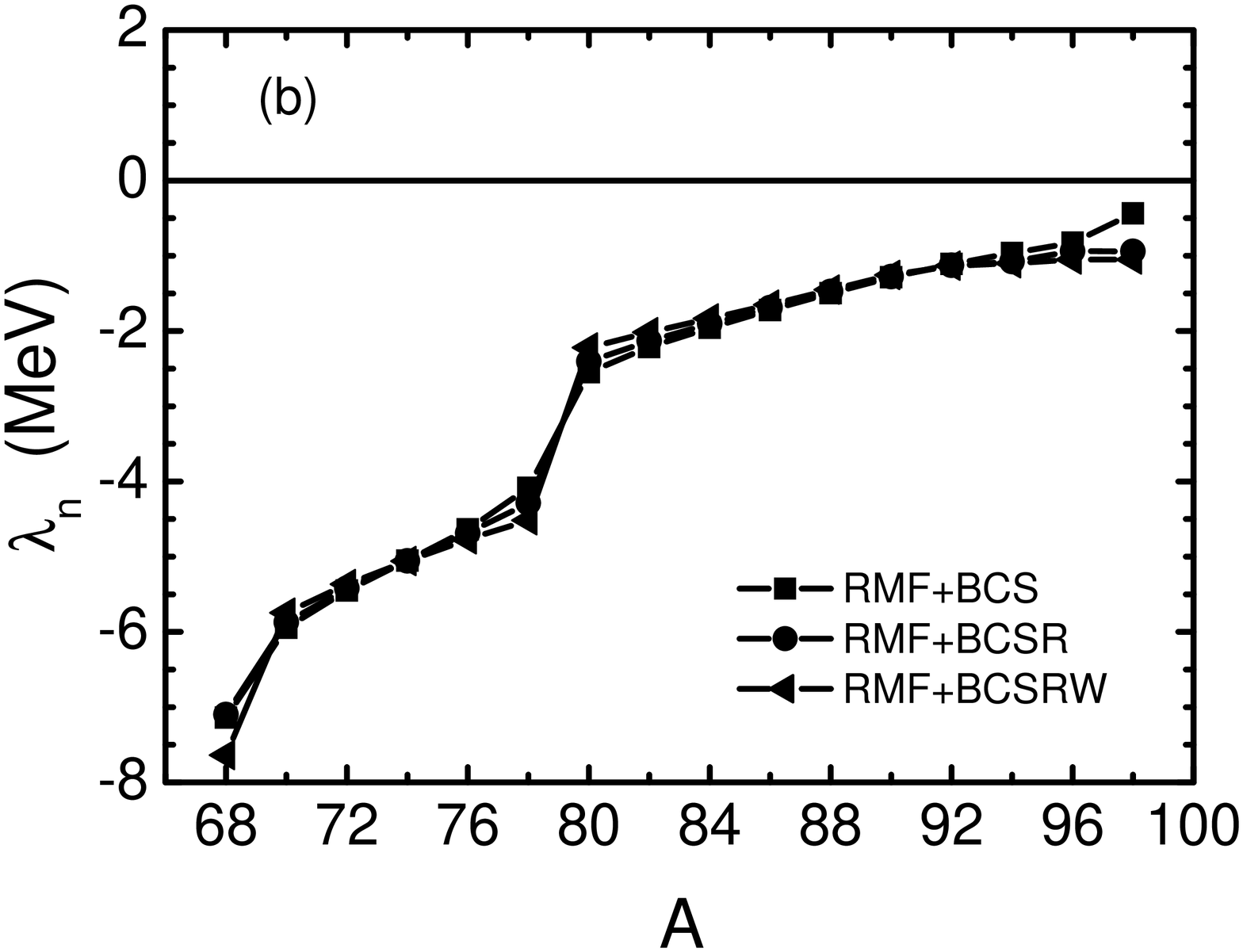}
\vglue -4.0cm \caption{The neutron pairing gap (a) and Fermi
energy (b) as a function of the atomic number $A$.  The notations
are same as in Fig.1.} \label{Fig2}
\end{figure}

\begin{figure}[tbp]
\includegraphics[scale=0.5,angle=0.]{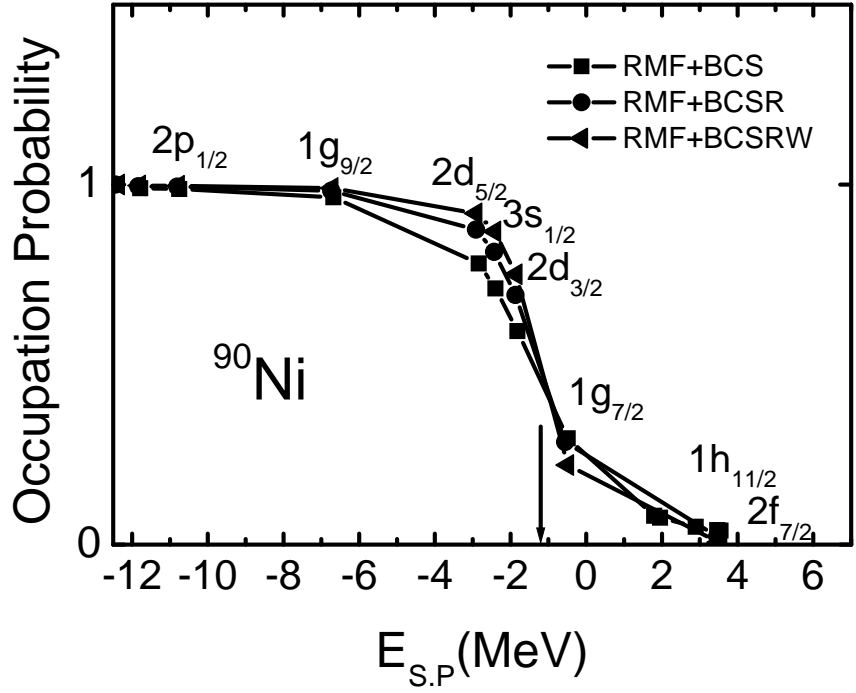}
\vglue -4.0cm \caption{The occupation probabilities for $^{90}$Ni
as a function of the single particle energy around the threshold.
The results are performed in RMF+BCS, RMF+BCSR and RMF+BCSRW
approaches with the effective nonlinear interaction NL3. The
notations are same as in Fig.1. } \label{Fig3}
\end{figure}

\begin{figure}[tbp]
\includegraphics[scale=0.5,angle=0.]{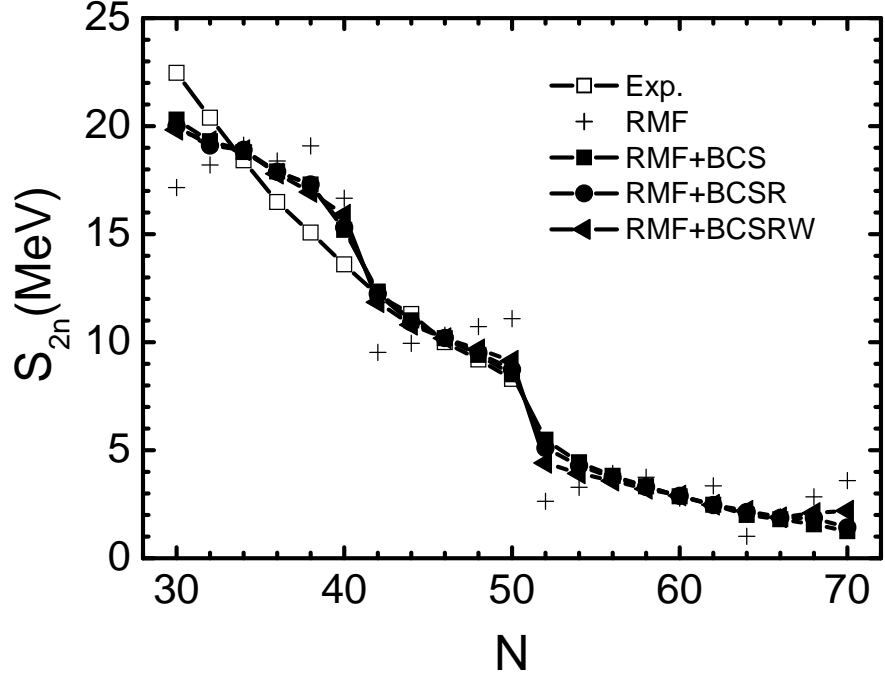}
\vglue -4.0cm \caption{Two-neutron separation energies for neutron
rich even-even Ni isotopes calculated in the RMF, RMF+BCS,
RMF+BCSR, and RMF+BCSRW approaches with parameter set NL3. The
cross signs are the results calculated in the RMF. The
experimental data denoted by open squares are taken from
Ref.\cite{Audi95}. Other notations are same as in Fig.1.}
\label{fig4}
\end{figure}

\begin{figure}[tbp]
\includegraphics[scale=0.5,angle=0.]{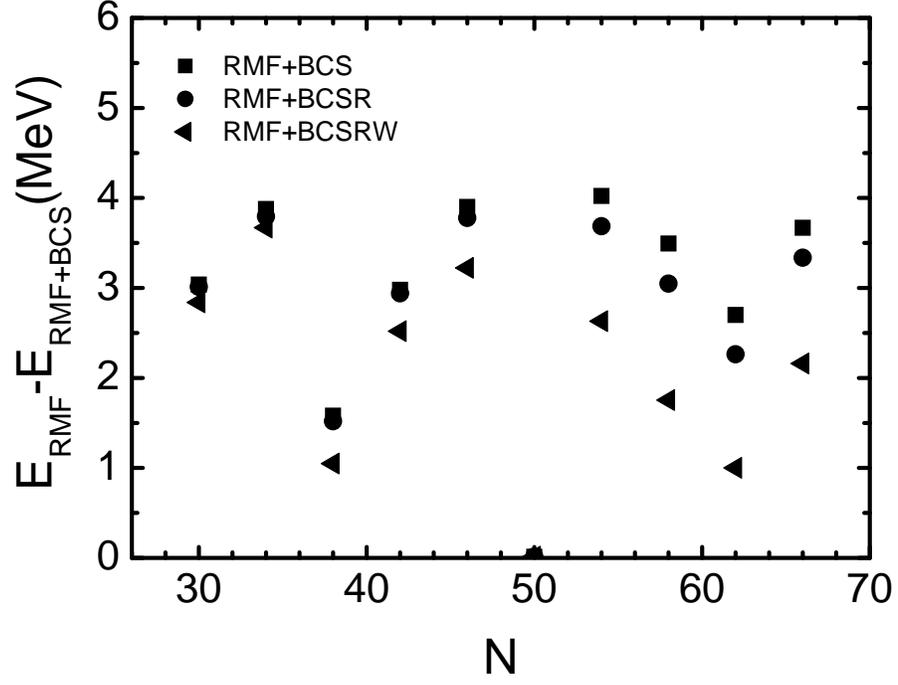}
\vglue -4.0cm \caption{The pairing correlation energies $E_{BCS}$
calculated in Eq.(35) in the RMF+BCS, RMF+BCSR and RMF+BCSRW
approaches. The notations are same as in Fig.1. } \label{fig5}
\end{figure}

\begin{figure}[tbp]
\includegraphics[scale=0.5,angle=0.]{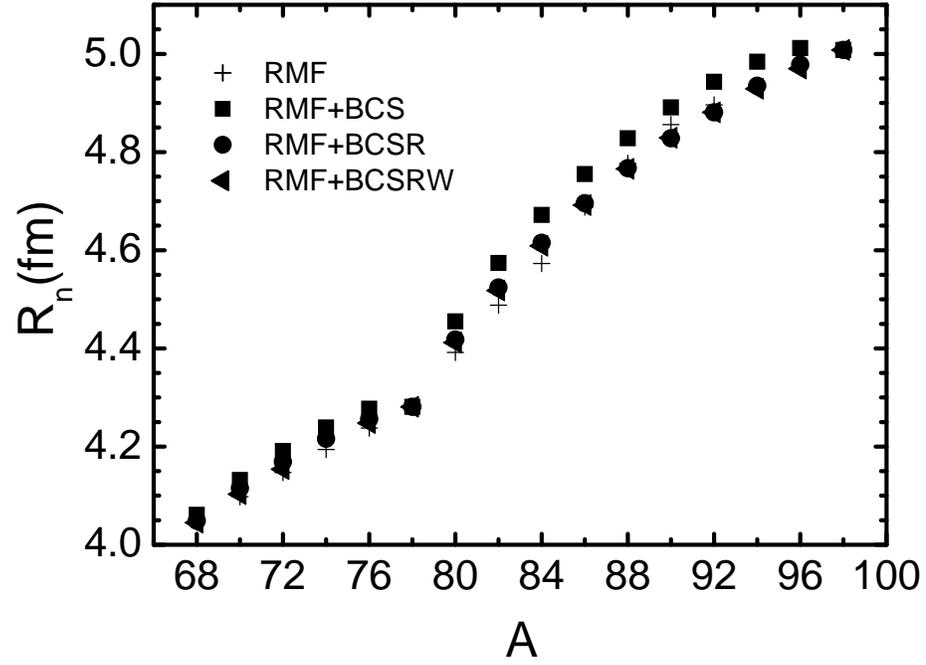}
\vglue -4.0cm \caption{Neutron rms radii for neutron rich
even-even Ni isotopes calculated in the RMF, RMF+BCS, RMF+BCSR,
and RMF+BCSRW approaches with parameter set NL3. The notations are
same as in Fig.4 } \label{fig6}
\end{figure}

\begin{figure}[tbp]
\includegraphics[scale=0.5,angle=0.]{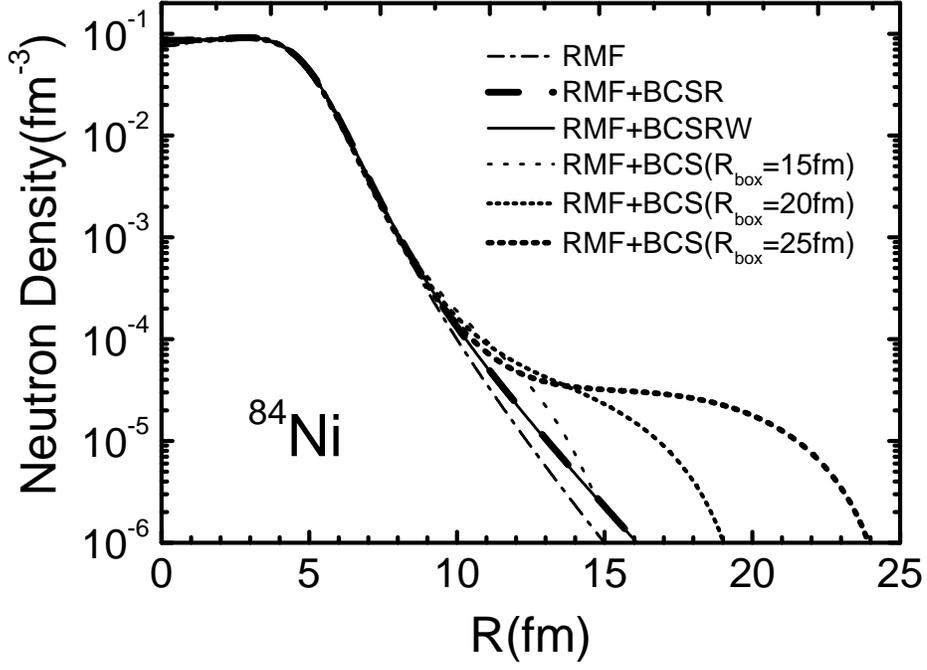}
\vglue -4.0cm \caption{Neutron density distribution for $^{84}$Ni
calculated in the RMF, RMF+BCS, RMF+BCSR, and RMF+BCSRW
approaches, where the RMF+BCS results calculated with various
values of the box size R$_{\text{box}}$ = 15, 20, 25 fm are also
plotted.} \label{fig7}
\end{figure}

\end{document}